%% file: main.tex
\documentclass[
	a4paper, 
	10pt, 
	unnumberedsections, 
	twoside, 
]{LTJournalArticle}

\makeatletter
\newcommand\notsotiny{\@setfontsize\notsotiny\@vipt\@viipt}
\makeatother

\runninghead{Improved RF Performance of Niobium Cavities via In-situ Vacuum Heat Treatment Technique. Y.Tamashevich et al.} 

\footertext{\textit{Preprint ver.6, Dec 2024 }} 

\usepackage{nameref}

\usepackage{siunitx}
\DeclareSIUnit\bar{bar}

\DeclareSIUnit{\atpercent}{at\%}

\usepackage[dvipsnames]{xcolor}

\usepackage[switch]{lineno} 

\usepackage{multirow}

\title{Improved RF Performance of Niobium Cavities via\\ \textit{In-situ} Vacuum Heat Treatment Technique}

\author{%
    Yegor Tamashevich\textsuperscript{1}
    ,
	Alena Prudnikava\textsuperscript{1,}\thanks{Corresponding author: \href{mailto:alena.prudnikava@helmholtz-berlin.de}{alena.prudnikava@helmholtz-berlin.de}}, 
	Aleksandr Matveenko\textsuperscript{1},\\
    Axel Neumann\textsuperscript{1},
    Oliver Kugeler\textsuperscript{1},
    Jens Knobloch\textsuperscript{1,2}\\ 
}

\date{\footnotesize\textsuperscript{
\textbf{1}}Helmholtz Centre for Materials and Energy, Albert-Einstein-Str. 15, 12489 Berlin, Germany\\
\textsuperscript{\textbf{2}}Department of Physics, Universität Siegen, Walter-Flex-Str. 3, 57068 Siegen, Germany
}

\renewcommand{\maketitlehookd}{%
	\begin{abstract}
		\noindent 
        \subfile{sections/0-Abstract}  
	\end{abstract}
}

\begin{document}


\maketitle 


\section{Introduction}
\label{introduction}
\subfile{sections/1-Introduction}

\section{Experimental}
\subsection{Setup}
\label{setup}
\subfile{sections/2-Setup}

\subsection{Procedure}
\label{procedure}
\subfile{sections/3-Procedure}

\section{Results}
\label{results}
\subfile{sections/4-Results}

\section{Summary and Conclusion}
\label{summary}
\subfile{sections/5-Summary}

\section{Acknowledgements}
\label{acknowledgements}
\subfile{sections/6-Acknowledgements}



\printbibliography 


\end{document}

%% file: sections/0-Abstract.tex

Vacuum thermal treatments (baking) are known to improve the superconducting properties of the RF surface layer of niobium cavities, and are employed as a last processing step to increase their efficiency determined by intrinsic quality factor Q$_0$. 
A new method to perform the baking has been demonstrated.
It consists in annealing of an evacuated cavity with the local heaters installed on its outer surface in a cryostat which ensures an exterior vacuum and protects the outer cavity surface from oxidation.
Such a set-up has a number of advantages as it does not require to cool the cavity flanges during baking, and allows to perform the "cold" RF characterization of the cavity \textit{in situ}, immediately after the thermal treatment without disassembly of heating elements.
Moreover, the air exposure that causes partial degradation of Q$_0$ by surface reoxidation is avoided.
The heat treatment of a single-cell \SI{1.3}{\giga\hertz} niobium cavity at \SI{230}{\degreeCelsius} for \SI{24}{\hour} demonstrated the doubling of Q$_0$ at $E_{acc}$=\SI{10}{\mega\volt\per\meter} (from \SI{1.20e10}{} to \SI{2.4e10}{}) and retained the maximal accelerating field of \SI{35}{\mega\volt\per\meter} without quenching.
The selection of treatment parameters is based on our previous XPS studies. 
This treatment ensures incomplete dissolution of the native oxide by oxygen diffusion, thereby preventing interaction of niobium surface with external contaminants.
We propose to bake the cavities directly in a cryomodule, which would allow to use the treatment to improve their performance.
The potential impact of material parameters on the components of surface resistance has been briefly examined.

%% file: sections/1-Introduction.tex
g radiofrequency (SRF) cavities form the core of modern accelerators and free-electron lasers, 
various fields of science, medicine, and industry.
The efficacy of SRF cavities hinges on their intrinsic quality factor, $Q_0$, and its dependence on the accelerating gradient, $E_{acc}$.
A high $Q_0$ signifies minimal heat dissipation and reduced cryogenic power requirements.
Since the efficiency of \SI{2}{\kelvin} cryogenic plants is in the per mil range, even small gains in $Q_0$ have a large impact on the AC power consumption and the complexity of the plant.

The intrinsic quality factor is inversely related to the resistance of a thin surface layer that interacts with the RF field during low-temperature cavity operation.
extensive research has been conducted worldwide to explore the effects of various thermal treatment methods on the superconducting properties of this surface layer.
Some methods involve simple vacuum annealing at different temperatures and durations (e.g., mild baking, medium-temperature baking \autocite{palmer1987influence, ciovati2006improved, visentin1999cavity, romanenko2013effect, posen2020ultralow,ito2021systematic}), while others incorporate gas environment with or without subsequent chemical processing (e.g., nitrogen doping and infusion \autocite{grassellino2013nitrogen, grassellino2017unprecedented, dangwal2018surface, prudnikava2022systematic}).
Regardless of the method, these treatments increase $Q_0$ by optimizing the concentration of point defects in niobium lattice within the RF penetration depth.

The improvement in $Q_0$ is attributed to several phenomena.
Firstly, this involves reducing the mean free path of electrons in the normal state, which decreases the temperature-dependent component of niobium resistance in the superconducting state (the Bardeen-Cooper-Schrieffer resistance, $R_{BCS}$) \autocite{padamsee2008rf}.
Additionally, flux pinning is enhanced, which reduces the residual surface resistance component,  $R_{res}$ \autocite{dhavale2012flux, gonnella2016impact}.
Point defects also act as trapping centers, inhibiting the formation of lossy niobium hydrides, which tend to form on the surface of niobium during cooling to low temperatures 
Recently, medium-temperature baking, often referred to as Mid-T baking, has garnered significant attentiomid-Ta simple method to improve the $Q_0$ of niobium cavities. 
The standard mid-T baking process involves heating the cavity inside a dedicated vacuum furnace.
However, this approach has several drawbacks that limit the cavity performance.
Although protective caps have been used in some experiments to mitigate this risk  especially if the protective Nb$_2$O$_5$ layer is completely dissolved \autocite{prudnikava2024situ}\autocite{ito2021systematic}, contamination remains a significant concern.
Moreover, post-treatment exposure to ambient air results in partial reoxidation, diminishing the benefits of the baking process \autocite{posen2020ultralow}.

To address these challenges, alternative approaches have been explored, such as sealing the cavity volume and connecting it to a separate pumping station during heat treatment \autocite{posen2020ultralow}.
This method requires protecting vacuum gaskets from thermal degradation by incorporating additional cooling equipment for the cavity flanges. 
Despite these advancements, the current mid-T baking procedure
remains costly and complex, requiring specialized equipment like clean vacuum furnaces and flange cooling systems, and time-consuming procedures.

In this work, we demonstrate a new, simplified method for vacuum annealing niobium cavities that improves their RF performance.
This method does not require a traditional vacuum furnace or flange cooling and avoids air exposure of the cavity prior to RF operation.

%% file: sections/2-Setup.tex

The heat treatment was conducted within the Small Vertical Test Stand (SVTS) cryostat at HZB.
The cryostat testing chamber has a diameter of \SI{600}{\milli\meter} and a height of \SI{2000}{\milli\meter}.
It is equipped with a helium level probe, multiple PT100 and Cernox temperature sensors, and a resistive heater for helium evaporation.
Additionally, a magnetic shield is installed inside the chamber.
None of these auxiliary components were dismounted for the experiment.

The heat-treatment method was applied to a standard \SI{1.3}{\giga\hertz} single-cell TESLA-shape niobium cavity (1DE13). 
The cavity (RRR=230) was previously treated with buffer chemical polishing, annealed at \SI{800}{\degreeCelsius}, followed by electropolishing, high-pressure rinsing (HPR) and a \SI{120}{\degreeCelsius} bake.
It was then stored in air for several years.
Prior to the experiment, it underwent HPR at \SI{100}{\bar} at HZB.
The cavity was equipped with a high-Q input antenna (attached to the top flange via a 7/16 feedthrough) and a field-probe antenna (connected to the bottom flange via an NW16 SMA feedthrough).
A vacuum right-angle valve was also attached to the bottom flange.
The cavity was placed on a titanium retaining plate within a standard SVTS insert, though it was not tightly connected to the insert.
A particle-free pumping line (CF40 bellow) 
was attached to the cavity right-angle valve.

The same SVTS insert was used for the initial baseline RF test, subsequent heat treatments, and follow-up RF tests.
Throughout all phases, the cavity vacuum remained better than \SI{1e-6}{\milli\bar}.
For the vertical cold RF test, the cavity was equipped with two Cernox sensors and two RF cables.

\subsection{Preparation for the Heat Treatment}
efore each heat treatment, following the cold test, the RF cables and the Cernox sensors were removed from the cavity and placed at the top part of the insert (though the cables and feedthroughs remained attached to the insert).
Several additional feedthroughs were installed on the insert, including a 4-pin high-voltage and high-current CF40-feedthrough, a KF-40 5-channel K-type thermocouple feedthrough, and a KF-40 5-channel E-type thermocouple feedthrough.

\begin{figure} 
	\includegraphics[width=\linewidth]{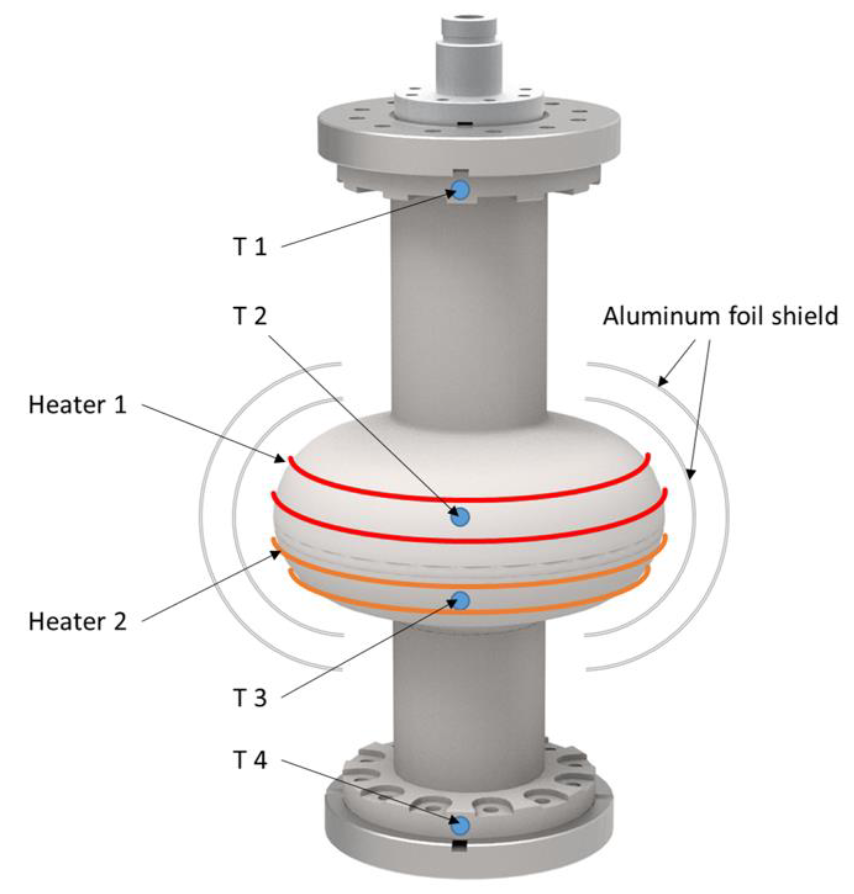}
	\caption{Schematic view of the experimental setup. T1, T4 are E-type termocouples at the cavity flanges; T2, T3 are E-type termocouples at cavity half-cells. Resistive heaters 1,2 are wrapped in aluminum foil. Two K-type control thermocouples are located between the heaters and the cavity surface (not shown). The cavity cell is wrapped in an aluminum-foil shield.}
	\label{fig:cavity_model}
\end{figure}

Four E-type thermocouples were installed  in the configuration shown in figure \ref{fig:cavity_model}.
Thermocouples T1 and T4 were attached to the cavity flanges, thermocouples T2 and T3 were affixed to the half-cells.
All thermocouples were secured with Kapton tape.

\begin{figure} 
	\includegraphics[width=\linewidth]{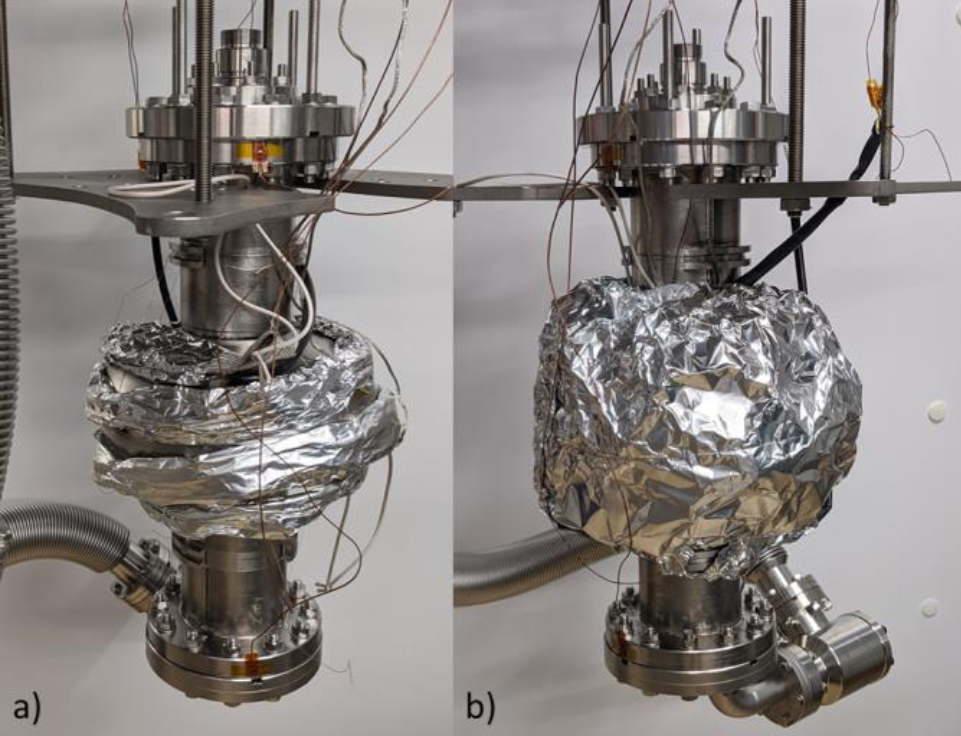}
	\caption{Image of the cavity with heaters wrapped in aluminum foil to avoid cavity contamination (a), and with external aluminum-foil thermal shield (b).}
	\label{fig:cavity_image}
\end{figure}

Two resistive heaters (Horst HBS010, \SI{250}{\watt} nominal power each) were connected to the high-voltage feedthrough allowing independent control.
The heaters were operated using two Horst HT31N temperature regulators.
To avoid potential contamination of the outer cavity surface, the heaters were wrapped with in aluminum foiler 1 was positioned around the top half-cell, while Heater 2 was placed around the bottom half-cell (see figure \ref{fig:cavity_model} and figure \ref{fig:cavity_image}(a)).
The heaters were fixed in place using copper wire.
Additionally, two K-type control thermocouples were positioned between the heaters and the cavity surface.

The cavity cell was further wrapped in aluminum foil, as shown in figure \ref{fig:cavity_image}(b) although
the beam tubes and flanges were left uncovered.

%% file: sections/3-Procedure.tex

After the preparation, the insert was installed in the cryostat.
The external particle-free pumping line connected to the insert was linked to the pumping station,
ensuring the cavity was actively pumped throughout the baking process.
The helium space within the cryostat was evacuated to a pressure of \SI{1e-5}{\milli\bar}, while the isolation vacuum space was vented with nitrogen to help cool the helium vessel.
Hence, both the interior and exterior of the cavity were maintained under vacuum conditions during baking.
Prior to heating, the pressure inside the cavity was \SI{9e-7}{\milli\bar}.

\subsection{Choosing the Parameters of Heat Treatment}
\begin{figure} 
	\includegraphics[width=\linewidth]{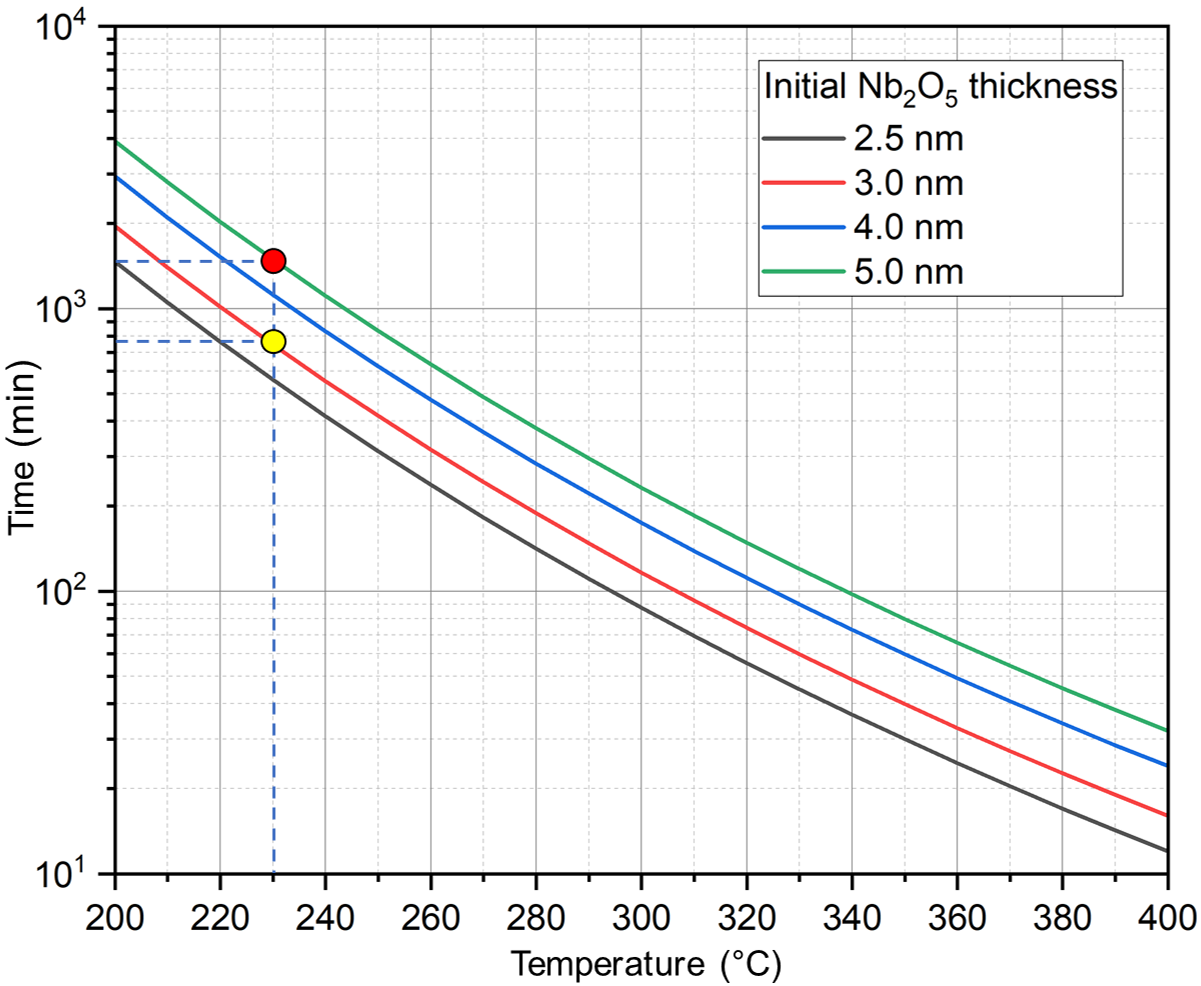}
	\caption{Recipe determined from \textit{in-situ} baking XPS studies \autocite{prudnikava2024situ} for the reduction of the Nb$_2$O$_5$ layer. The curves denote the required baking time as a function of temperature to reduce the initial Nb$_2$O$_5$ layer to \SI{1}{\nano\meter} thickness. The cavity studied for this work had an oxide thickness of about \SI{5}{\nano\meter}.}
	\label{fig:Figure 1 nm_v3.png}
\end{figure}

The selection of heat treatment parameters was guided by results from our previous work \autocite{prudnikava2024situ}.
By studying the kinetics of niobium oxide dissolution \textit{in situ} at various temperatures using  synchrotron X-ray photoelectron spectroscopy, it has been determined that a ~\SI{1}{\nano\meter}-thick Nb$_2$O$_5$ layer substantially reduces interaction with surface contaminants, such as carbon.
Additionally, the doping of the niobium surface layer occurs in a controlled manner, driven by oxygen originating from the native oxide reduction process. 
Figure \ref{fig:Figure 1 nm_v3.png} illustrates the temperatures and durations required for the heat treatment to satisfy these conditions for different initial Nb$_2$O$_5$-layer thicknesses, which are typical for chemically treated niobium cavities.

To select appropriate parameters, the temperature and duration of baking should fall below the curve in figure \ref{fig:Figure 1 nm_v3.png} for the corresponding initial oxide thickness.
For the first heat treatment, a low temperature range was chosen due to uncertainties regarding the heat load on the flanges and the required heating power.
A lower baking temperature also reduces the oxide reduction rate, simplifies control of experimental parameters, and lowers the risk of failure.
The goal was to keep the vacuum gaskets of the cavity ports below \SI{300}{\degreeCelsius}, with a significant margin (approximately \SI{50}{\degreeCelsius}).
therefore, the temperature range selected for baking was \SIrange{230}{240}{\degreeCelsius}.

The heat treatment has been performed in two steps, with an RF test in between each step. 
The objective was to leave about \SI{1}{\nano\meter} of niobium pentoxide after the second treatment.
Since the cavity had been stored in an open environment for an extended period, we assumed that the Nb$_2$O$_5$ layer was at least \SI{5}{\nano\meter} thick, which is greater than the \SIrange{2.7}{3.5}{\nano\meter} thickness of freshly prepared niobium used in \autocite{prudnikava2024situ} for heat treatment studies.
The plan was to dissolve approximately \SI{2.2}{\nano\meter} of Nb$_2$O$_5$ during the first treatment and \SI{1.8}{\nano\meter} during the second.
According to \autocite{prudnikava2024situ} at \SI{230}{\degreeCelsius}, this would require about \SI{13}{\hour} and \SI{11}{\hour}, respectively.

\begin{figure} 
	\includegraphics[width=\linewidth]{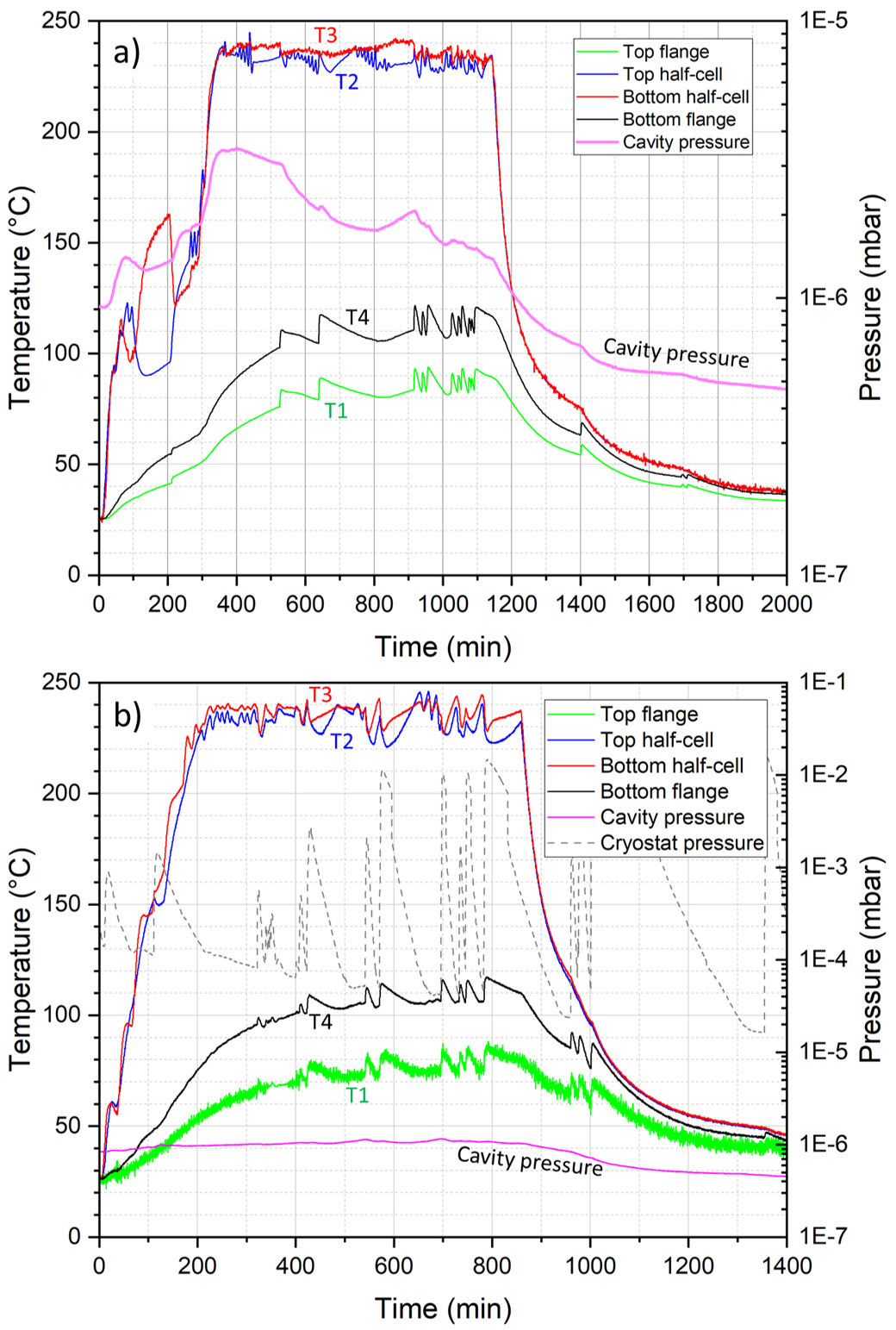}
	\caption{Temperature of flanges and a top and a bottom half-cells of the cavity, as well as pressure in the cavity during the first (a) and the second (b) heat treatments. For the second heat treatment the cryostat pressure is shown.} 
	\label{fig:temperature_plot}
\end{figure}

Figure \ref{fig:temperature_plot} shows the temperature and pressure plots during the first (a) and the second (b) heat treatments of the cavity in the cryostat. 
Several vacuum events were observed during the heating process, with the pressure in the evacuated helium space increasing from approximately \SI{1e-5}{\milli\bar} to \SI{1e-2}{\milli\bar}.
At this higher pressure, heat transfer by gas is non-negligible as evidenced by the thermocouple readings attached to the flanges.
These vacuum events also occurred before and after heating, potentially caused by virtual leaks from the heaters and cables, which are likely not compatible with high-vacuum environment.

During the experiment, the bottom flange temperature remained below \SI{120}{\degreeCelsius} while the top flange stayed below \SI{90}{\degreeCelsius}.
As shown in figures \ref{fig:cavity_image}(a,b), the top flange rests on the holding plate.
The contact between the flange and the plate is made via bolts that extend beyond the flange.
Due to the relatively small contact area, we assume that the lower temperature of the top flange results from radiation shielding by the holding plate rather than contact cooling.

The total time-averaged heating power during this experiment ranged between \SIrange{50}{80}{\watt}, with both heaters operating at duty cycles between \SI{10}{\percent} and \SI{15}{\percent}.

%% file: sections/4-Results.tex
\begin{figure} 
	\includegraphics[width=\linewidth]{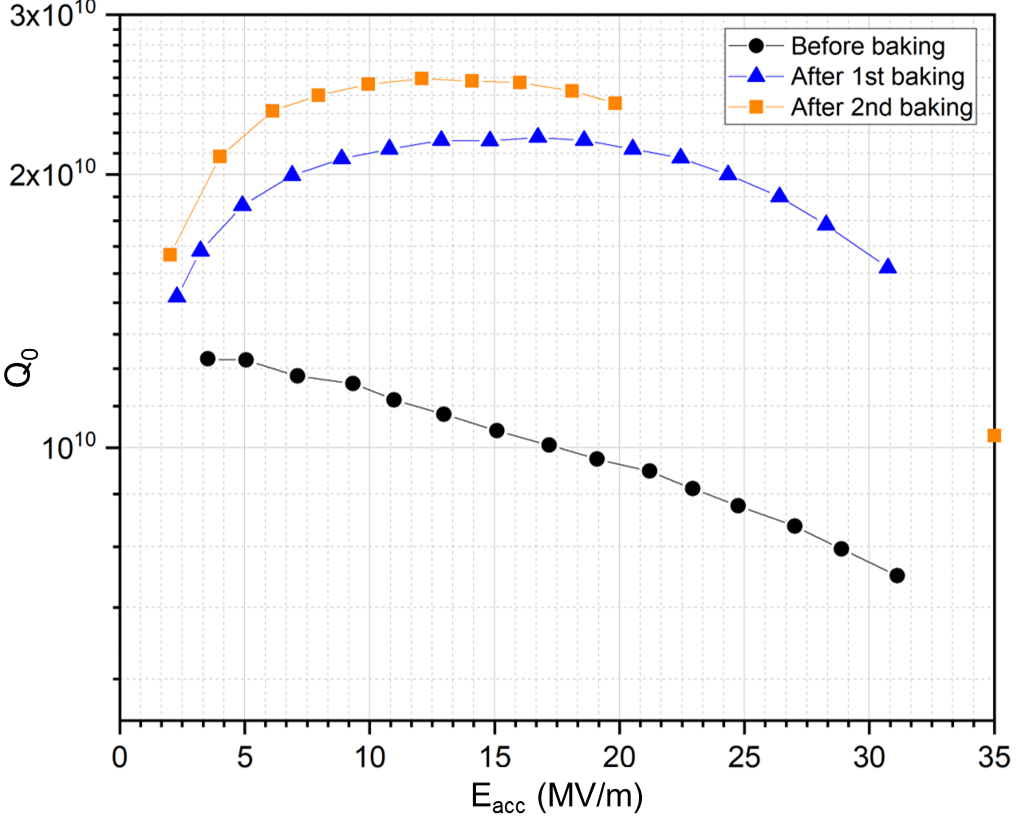}
	\caption{$Q_0$($E_{acc}$) plots before and after the respective heat treatment ($T$=\SI{1.8}{\kelvin}, $p_{He}$=\SI{16}{\milli\bar}).     
     After the second baking, no points were measured between \SI{20}{\mega\volt\per\meter} and \SI{35}{\mega\volt\per\meter} due to an administrative limit imposed to ensure no $Q_0$ degradation takes place in case of a quench. Only at the very end of the test was the data point at \SI{35}{\mega\volt\per\meter} obtained.
    }
	\label{fig:Figure QvsE_v1}
\end{figure}

The cavity was tested vertically before and after each heat treatment.
Throughout the experiment, the cavity remained mounted in the insert and under vacuum.
The plots of the intrinsic quality factor ($Q_0$) versus the accelerating field ($E_{acc}$) were measured at 1.51, 1.64, 1.72, 1.80, 1.86, 1.93, 1.99 and \SI{2.04}{\kelvin}. The $Q_0$($E_{acc}$) curves measured at $T$=\SI{1.80}{\kelvin}, before and after heat treatments, are shown in figure \ref{fig:Figure QvsE_v1}.

Before the heat treatments, the cavity performance was moderate.
At $T$=\SI{1.80}{\kelvin} ($p_{He}$=\SI{16}{\milli\bar}), it had a maximum quality factor of \SI{1.2e10}{} at $E_{acc}$=\SI{5}{\mega\volt\per\meter}.
The previous cold test in 2018 showed a higher $Q_0$, but since the cavity had been stored open to air afterward, it is assumed that this degradation was caused by heavy oxidation.

The onset of field emission occurred at approximately $E_{acc}$=\SI{25}{\mega\volt\per\meter}.
The field emission was unstable and could be processed away, allowing the final point at \SI{31}{\mega\volt\per\meter} to be measured without field emission (radiation values are not shown in the plots).
However, field emission reappeared during the subsequent measurement at around \SI{25}{\mega\volt\per\meter}.
We chose not to process the field emission at fields above \SI{25}{\mega\volt\per\meter} as this could degrade $Q_0$ or lower the field emission onset.
As a result, an administrative limit of $E_{acc}$=\SI{31}{\mega\volt\per\meter} was set for all tests.

After the first heat treatment, the $Q_0$ of the cavity increased from \SI{1.20e10}{} to \SI{1.75e10}{} at $E_{acc}$=\SI{5}{\mega\volt\per\meter} (figure \ref{fig:Figure QvsE_v1}).
Additionally, an anti-Q-slope effect --- i.e., continuous improvement in the quality factor at moderate field levels --- was observed \autocite{martinello2018field}.
At \SI{1.80}{\kelvin}, the maximum quality factor of \SI{2.2e10}{} was achieved at $E_{acc}$=\SI{17}{\mega\volt\per\meter}. 
The onset of field emission remained at about $E_{acc}$=\SI{25}{\mega\volt\per\meter}, and, as in the previous test, the field emission could be processed away before reaching the maximum field.
The tests were again carried out to the administrative limit of $E_{acc}$=\SI{31}{\mega\volt\per\meter}.

After the second heat treatment, $Q_0$ increased further to \SI{2.20e10}{} at $E_{acc}$=\SI{5}{\mega\volt\per\meter} with the maximum $Q_0$ shifting to lower fields, reaching \SI{2.45e10}{} at $E_{acc}$=\SI{13}{\mega\volt\per\meter}.
The cold test was routinely performed at fields below \SI{20}{\mega\volt\per\meter} due to administrative limits of SRF facility at that dates.
The point at \SI{35}{\mega\volt\per\meter} was measured at \SI{1.8}{\kelvin} only to prove no cavity degradation.

\begin{figure} 
	\includegraphics[width=\linewidth]{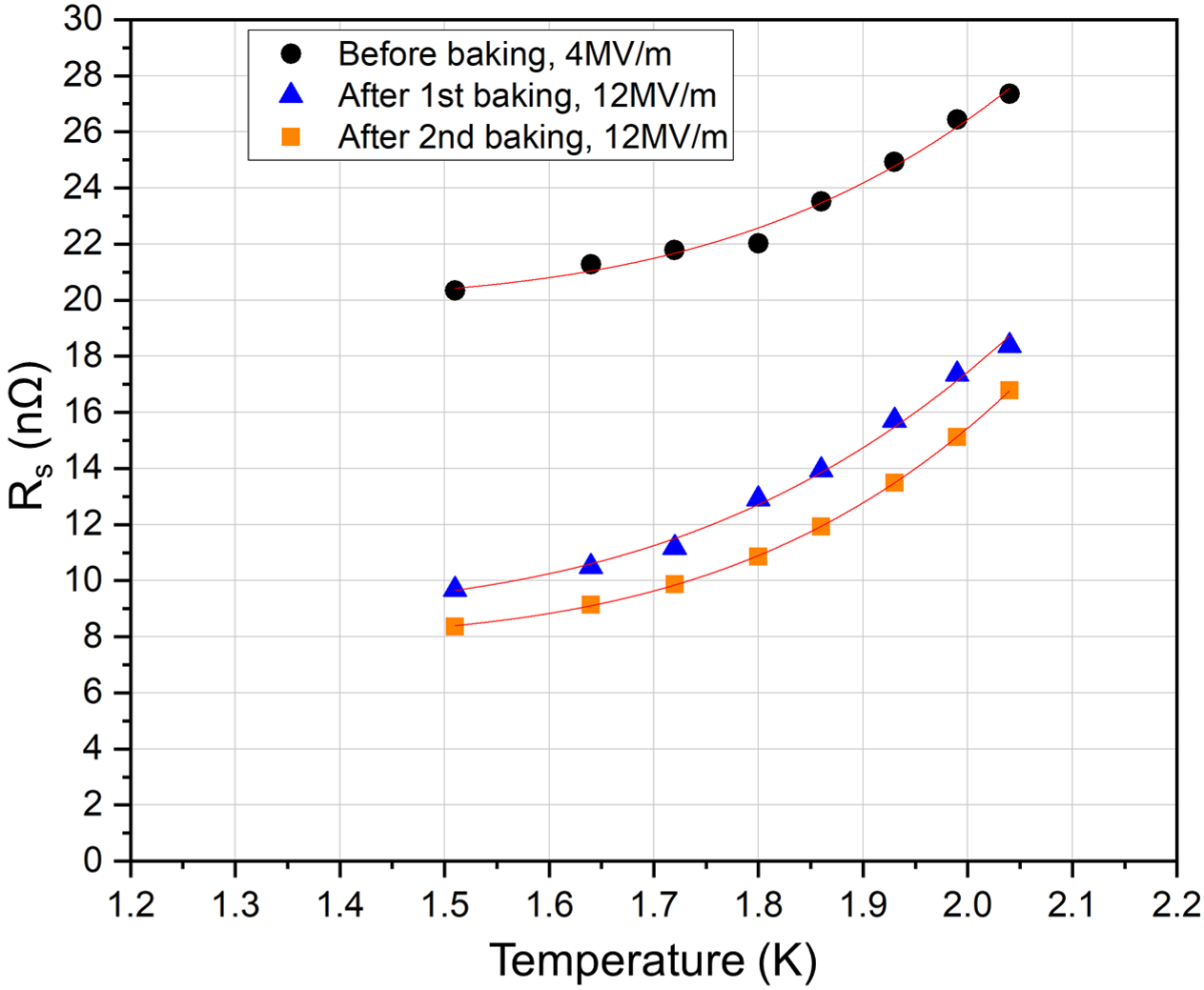}
	\caption{Temperature dependence of the surface resistance before and after the respective heat treatment. The red curves are the BCS fits according to Equation \ref{eq:surface_resistance} with varied $A$ (see text) and $\Delta/k_BT_c$ = 1.89.}
	\label{fig:Figure_RsvsT_v1}
\end{figure}

The surface resistance ($R_s$) versus temperature plots, along with the corresponding BCS fits (solid lines), are presented in figure \ref{fig:Figure_RsvsT_v1}.
The following function was used for the fitting \autocite{padamsee2008rf}:

\begin{equation}
R_s=A\frac{f^2}{T}e^{\frac{-\Delta}{kT}}+R_{res},
\label{eq:surface_resistance}
\end{equation}

where $R_s$ is the surface resistance, $A$ is a fitting parameter related to the BCS resistance, $T$ is the temperature, $f$ is the frequency of the cavity, $k$ is the Boltzmann constant, $\Delta$ is the superconducting energy gap, and $R_{res}$ is the residual resistance.

The best fits were obtained with a superconducting energy gap, $\Delta$, of \SI{1.50}{\milli\electronvolt}.
The ratio $\Delta/k_BT_c$=1.89 (with $T_c$=\SI{9.208}{K}) is consistent with the values reported for nitrogen-doped (N-doped) cavities \autocite{maniscalco2017importance}.
The residual resistance ($R_{res}$) before baking was \SI{19.83\pm0.19}{\nano\ohm} at $E_{acc}$=\SI{4}{\mega\volt\per\meter}, and after the first and second baking steps, it dropped to \SI{9.11\pm0.19}{\nano\ohm} and \SI{7.68\pm0.05}{\nano\ohm} (both at $E_{acc}$=\SI{12}{\mega\volt\per\meter}), respectively.

The fitting parameter A was:
\begin{itemize}
    \item \SI{4.70\pm0.21e-23}{\ohm\kelvin\per\hertz\squared} before baking,
    \item \SI{5.95\pm0.21e-23}{\ohm\kelvin\per\hertz\squared} after the first baking and
    \item \SI{5.52\pm0.05e-23}{\ohm\kelvin\per\hertz\squared} after the second baking.
\end{itemize}

\begin{figure*} 
	\includegraphics[width=\linewidth]{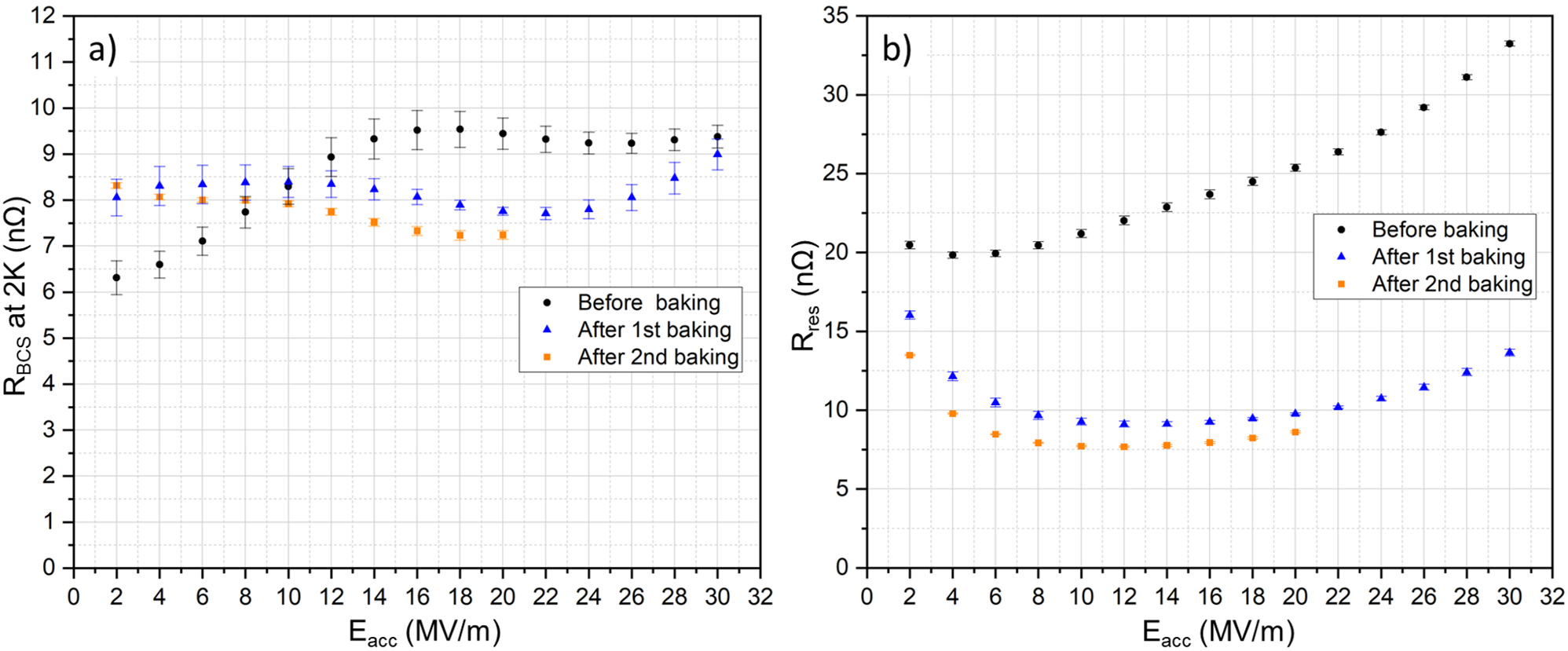}
	\caption{BCS resistance, $R_{BCS}$, (a) and residual resistance, $R_{res}$; (b) plots before and after the respective heat treatment ($T$=\SI{2}{\kelvin}, $p_{He}$=\SI{30}{\milli\bar}).}
	\label{fig:Figure_R_res}
\end{figure*}

The BCS surface resistance and residual resistance as functions of $E_{acc}$ at T=\SI{2}{\kelvin} are shown in figure \ref{fig:Figure_R_res}.
Before baking, $R_{BCS}$ increased with increasing field, which is typical for standard niobium cavities \autocite{ito2021systematic}.
However, after the heat treatments the $R_{BCS}$ curves became almost independent of the accelerating field.

As to $R_{res}$, it initially increased significantly with the field starting from \SI{6}{\mega\volt\per\meter} for the unbaked cavity.
After the baking, the dependence changed, i.e. $R_{res}$ decreased with the field up to $E_{acc}$=\SI{12}{\mega\volt\per\meter}, followed by a slow increase at higher fields.

\subsection{Relation of Surface Defects to Superconducting Surface Resistance}

It is generally accepted that vacuum heat treatment of niobium leads to the reduction of the native oxide layer, primarily due to the diffusion of oxygen into the near-surface region \autocite{delheusy2008x, king1990kinetic, ma2003angle, ciovati2006improved, veit2019oxygen, lechner2024oxide}.
As a result, the treated surface consists of a distorted lattice structure (following oxide dissolution), with lower-valence oxides, as well as a subsurface layer doped with oxygen as depicted schematically in figure \ref{fig:Schematic Surface layer_v1.png}.
To optimize thermal treatment parameters for cavity performance, it is crucial to analyze how these surface characteristics are affected by different heat treatments, as well as their correlation with superconducting surface resistance.

\subsubsection{BCS resistance and the mean-free path of "normal" electrons.}
\hfill\\ 
The $R_{BCS}$ component of the surface resistance is known to be related to the mean-free path of "normal" electrons \autocite{padamsee2008rf}.
The mean free path, in turn, is inversely proportional to the concentration of defects within the London penetration depth.
At low temperatures, the primary scattering centers in well-annealed niobium are interstitial atoms, particularly oxygen.
Thus, the oxygen concentration and its distribution play a significant role for cavity performance.

Using the solution of Fick’s second law for a continuous plane source \autocite{jaeger1959conduction} the oxygen-concentration depth profiles in niobium were calculated according to:

\begin{equation}
	C = \frac{1}{\sqrt{\pi D}}\int_{0}^{t}e^{-x^2/4Dt}\cdot\frac{\phi(t)}{\sqrt{t}}dt, 
	\label{eq:plane_source}
\end{equation}

where $D$ is the diffusion coefficient, $t$ is the duration of baking, $\phi(t)$ is the rate at which the diffusing species are introduced from the surface plane into the solid, i.e. the rate of oxide reduction, for the first-order reaction equal to $\phi(t)$=$k$$C_0$exp($-kt$). The rate constant for oxide reduction, obtained from our previous study  was used, $k$=33.96exp(-63.62/$RT$) \SI{}{\kilo\joule\per\mole} \autocite{prudnikava2024situ}, along with the diffusion coefficient $D_0$=\SI{1.53e-14}{\centi\metre\squared\per\second}, from \autocite{pick1982depth}.
The experimentally determined oxygen concentration in niobium oxides and interstitials from \autocite{prudnikava2024situ}, $C_0$=151 (the number of oxygen atoms relative to 100 niobium atoms averaged across several chemically treated niobium samples), was also applied.

\begin{figure} 
	\includegraphics[width=\linewidth]{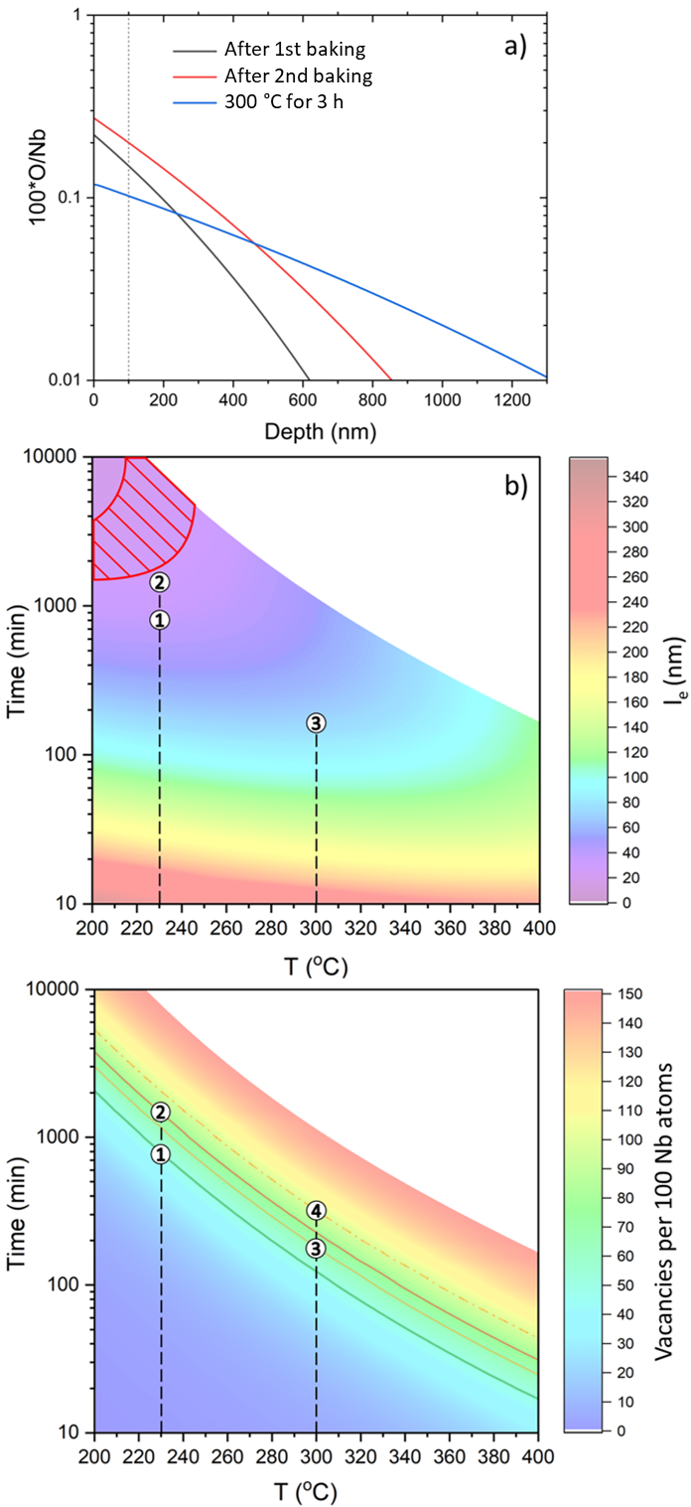}
	\caption{Theoretical modelling results: (a) the oxygen concentration depth profiles for various thermal treatments obtained using eq.(2); (b) the calculated mean free path of "normal" electrons, $l_e$, at the surface oxide layer/niobium interface. The red dashed area corresponds to the optimum (in terms of $R_{BCS}$) with $l_e$=20-30 nm. (c) The number of defects (oxygen vacancies) created in the top surface layer after the baking. Markers (1) and (2) refer to the calculated values of the first and second baking at \SI{230}{\degreeCelsius}, markers (3) and (4) refer to the calculated and experimental data for baking at \SI{300}{\degreeCelsius}/\SI{3}{\hour}, respectively. Solid lines represent contours of constant vacancy concentration, corresponding to the indicated concentration values.} 
	\label{fig:moddeling_results}
\end{figure}

The calculated concentration profiles of oxygen under the surface oxides for both the first and second heat treatments are shown in figure \ref{fig:moddeling_results}a.
For comparison, the oxygen profile from the mid-T baking conducted at \SI{300}{\degreeCelsius} for \SI{3}{\hour}, an optimal condition that has led to maximal $Q_0$ values in several studies, has been included. 
The results demonstrate that the average oxygen concentration within within the RF layer (ca. \SI{100}{\nano\meter}) increases from the first to the second heat treatment.
Furthermore, this value for both treatments exceeds that of the \SI{300}{\degreeCelsius}/\SI{3}{\hour}-baking process.

\begin{figure} 
	\includegraphics[width=\linewidth]{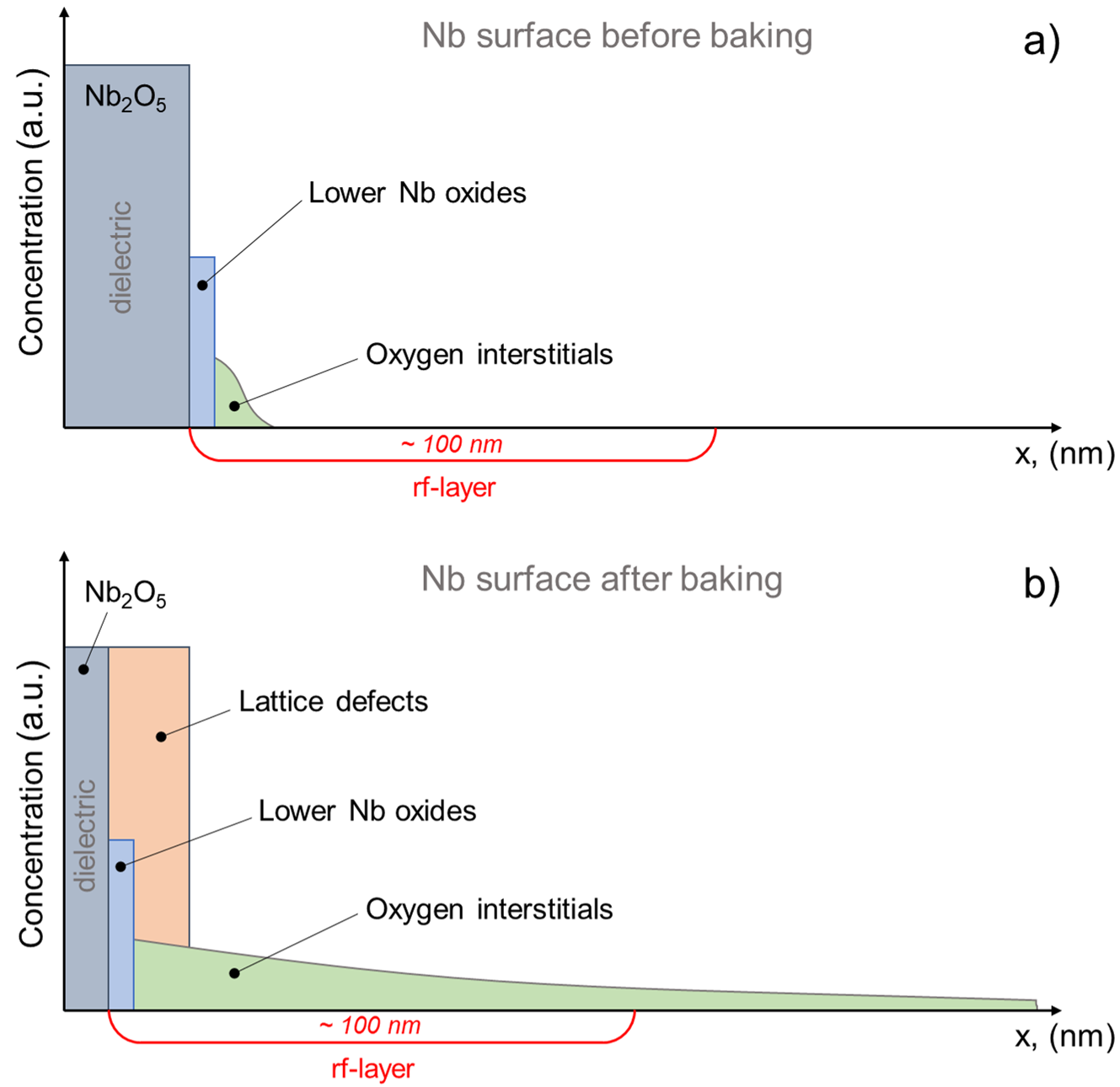}
	\caption{Schematic of the surface of Nb cavities (a) before and (b) after \textit{in-situ} mid-T baking (Note: the axes are not to scale).} 
	\label{fig:Schematic Surface layer_v1.png}
\end{figure}

The mean free path, $l_{e}$, was calculated using the known relation of $l_{e}$ to resistivity and the contribution of oxygen interstitials to resistivity \autocite{mayadas1972electrical, schulze1981preparation}. 
Figure \ref{fig:moddeling_results}b illustrates how $l_{e}$ varies with treatment duration and temperature, based on surface oxygen concentration under the oxide layer.
The minimal $R_{BCS}$ is predicted at $l_{e}$=\SIrange{20}{25}{\nano\meter} (red dashed area in figure \ref{fig:moddeling_results}b) \autocite{
maniscalco2017importance}.
After the first heat treatment at \SI{230}{\degreeCelsius}, the near-surface $l_{e}$ was around \SI{37.5}{\nano\meter} (marker (1) in figure \ref{fig:moddeling_results}b). 
The second heat treatment shifted $l_{e}$ closer to the theoretical optimum (\SI{30.5}{\nano\meter}, marker (2)), consistent with the observed reduction in $R_{BCS}$.
The mean free path corresponding to the \SI{300}{\degreeCelsius}/\SI{3}{\hour} treatment is shown by marker (3) in figure \ref{fig:moddeling_results}b).
One should note that the current diffusion model does not account for oxide reduction during the heating stage, which  plays a role in the \SI{300}{\degreeCelsius}/\SI{3}{\hour} treatment. 
Therefore, the actual oxygen profile and $l_{e}$=\SI{70.1}{\nano\meter} for the \SI{300}{\degreeCelsius}/\SI{3}{\hour} may differ slightly from the calculated values.
The XPS experimental data presented in \autocite{prudnikava2024situ} confirm a higher total oxygen concentration than predicted by the model.

As $l_{e}$ values are lower for the heat treatments at \SI{230}{\degreeCelsius} for a total time of \SI{13}{\hour} and \SI{24}{\hour}, it should result in lower $R_{BCS}$ compared to the \SI{300}{\degreeCelsius}/\SI{3}{\hour} treatment.
However, according to the obtained data, neither treatment achieved optimal $l_{e}$ values.
Notably, the oxygen concentration profile at \SI{300}{\degreeCelsius} is less steep compared to that at \SI{230}{\degreeCelsius} (figure \ref{fig:moddeling_results}a).
To our best knowledge, impact of this on $R_{BCS}$ remains unclear and requires further exploration.

Comparing the $R_{BCS}$($E_{acc}$) curves (figure \ref{fig:Figure_R_res}a) after the first and second baking and the initial cavity state, $R_{BCS}$ increases upon baking for fields up to \SI{8}{\mega\volt\per\meter}, and lower at moderate and high fields, with a minimum at \SIrange{18}{20}{\mega\volt\per\meter}, which would fit well for CW applications.
In other words, $R_{BCS}$($E_{acc}$) curves are more linear, i.e. less dependent on accelerating field.
The observed behavior requires further investigation with more cavities, as well as theoretical validation.
near-surface defects.
These defects are formed during the oxide reduction process when the native oxide layer is transformed into lower-valence oxides (non-superconducting at \SI{2}{\kelvin}) and distorted niobium latSuch(figure \ref{fig:Schematic Surface layer_v1.png}).
 tan-colored region in b Additionally, under certain baking conditions, niobium carbides may also form on the surface.It is known that normal conducting precipitates may act as strong pinning centers via condensation energy interaction \autocite{matsushita2000flux, matsushita2022flux}.  
Along with that, oxygen released into the lattice as interstitials, may alter the local elastic properties of the lattice, thereby affecting flux pinning behavior by elastic interactions.
Figure \ref{fig:moddeling_results}c shows the total number of oxygen atoms released from the oxide layer, which roughly corresponds to the number of oxygen vacancies created in the top surface layer (this number neglects formation of oxides with lower valences).
Owing to the model limitation mentioned above, the \SI{300}{\degreeCelsius}/\SI{3}{\hour} treatment shows a noticeable difference between theoretical (marker (3)) and experimental (marker (4)), data from \autocite{prudnikava2024situ}, with the theoretical data predicting a lower number of vacancies.
The values for the \SI{230}{\degreeCelsius} treatments are consistent with the XPS data obtained in our previous work \autocite{prudnikava2024situ}.

Upon baking, the $R_{res}$ component decreased significantly after the first heat treatment and showed a slight further reduction after the second treatment.
The number of vacancies in the top surface layer created by the first treatment was 49 per 100 Nb atoms, and this increased by 1.6 times after the second treatment.
Similarly, the number of oxygen interstitials within a \SI{100}{\nano\meter} depth was 18.3 per 100 Nb atoms and increased by a factor of 1.3.
Thus, the majority of defects are concentrated within a thin surface layer where the oxide was initially present.
We propose that, similar to the optimal value of $l_e$ that corresponds to minimal $R_{BCS}$, there exists an optimal number of defects within the top surface layer after oxide reduction, corresponding to the minimum $R_{res}$.
This hypothesis (described in \autocite{tamashevich2023model}) has yet to be verified by further experiments on cavities.

In the future, the quantity and dimensions of the lower-valence oxides created will be analyzed, as they are also considered efficient pinning centers.
It has been previously established that the contribution of defects of different dimensions varies with the applied field\autocite{pautrat2004quantitative, antoine2019influence}. 
Therefore,  since $R_{res}$ is not linearly related to the accelerating field, one can suggest that the size of defects affects the pinning behavior.

Estimating $R_{res}$ after air oxidation, which would “heal” all oxygen vacancies in the surface layer by transforming lower oxides into the highest oxidation state, while keeping the depth profile of interstices within the RF penetration depth, would provide useful information on the impact of surface defects on pinning.
The relationship between pinning effectiveness, surface defects (e.g., point defects, oxide layers or islands) and their properties (electronic, magnetic), and other surface properties like roughness, requires further study to improve our understanding of pinning mechanisms and $R_{res}$ behavior of baked niobium surfaces.

%% file: sections/5-Summary.tex

We have successfully demonstrated the feasibility of performing mid-T baking in-cryostat on superconducting cavities without requiring specialized vacuum furnaces.
Instead, a vertical cryostat was used as a vacuum chamber, presenting several advantages for research and development experiments in SRF science:

1) \textit{Preservation of cavity cleanliness:}
A key advantage is that the cavity's inner surface is never exposed to the vacuum environment of a traditional furnace.
This reduces the risk of contamination from furnace cleanliness issues and post-treatment reassembling, improving the likelihood of successful heat treatments when "safe" heat treatment parameters from \autocite{prudnikava2024situ} are applied to avoid carbide formation on the cavity surface.

2) \textit{Simplified flange cooling:} 
There is no need for separate pumping systems or active cooling of the flanges to protect gaskets, as temperatures at the flanges remain below the maximum allowable threshold for gaskets and feedthroughs, reducing the complexity of the procedure.

3) \textit{Enhanced post-treatment testing:} 
Immediate RF testing of the cavity following the baking process can be realized, without exposing the outer surface to air, hence preventing re-oxidation.
This may enhance the heat transfer from the cavity to the surrounding liquid helium, in which case it is expected that the cavity performance is improved.
Additionally, using non-magnetic heaters (e.g., tungsten, molybdenum, or ceramic) prevents magnetic flux trapping during cooldown.
These heaters can remain on the cavity during rf testing, streamlining the workflow.

4) \textit{Selective heating capability:} 
The ability to selectively heat specific regions of the cavity allows for greater experimental flexibility and the potential for optimized heat treatment parameters.

5) \textit{Retention of post-baking $Q_0$:} Avoiding air exposure between baking and RF operation fully preserves the $Q_0$ values achieved during treatment, preventing degradation due to reoxidation of the inner cavity surface \autocite{lechner2021rf}.

6) \textit{Application in real accelerators:}
The same procedure can potentially be adapted for the so-called "dressed" cavities installed in cryomodules, with heat treatment performed after installation \textit{in situ} using simple heaters, maximizing the benefits of mid-T baking.
This could fully harness the benefits of moderate-temperature baking in operational accelerators.
The SRF module with heaters and its manufacturing method are patent pending.

7) \textit{Optimized heat-treatment parameters:}
It was demonstrated that carefully choosing heat-treatment parameters (temperature and duration) that allow partial reduction of the native oxide layer through diffusion reliably improves the $Q_0$ of the cavity.

8) \textit{Surface resistance modulation:}
The relation of the change to the state of niobium surface characterized mainly by defects generation
in a thin upper layer and alteration of the mean free path of "normal" electrons within the RF penetration depth, to the surface resistance components modulation
has been analyzed. The future perspectives to
further improve $Q_0$ have been briefly discussed.

We consider this innovative approach to mid-T heat treatment represents a significant advancement in cavity treatment methodologies.
This method offers a practical and versatile solution for accelerating progress in SRF science while enabling performance improvements in real-world accelerator systems.
A notable advancement is the successful mid-T baking of a single-cell niobium cavity without the use of a conventional furnace and venting the cavity between heat treatment and testing.
This approach not only streamlines the process by reducing equipment costs but also simplifies experimental procedures.

%% file: sections/6-Acknowledgements.tex

We acknowledge Jan Ullrich, Henry Plötz, Michael Schuster, and Sascha Klauke for clean-room work and engineering. We axpress our thanks to our cryogenic team, especially Axel Hellwig, Karsten Janke, and Stefan Rotterdam.